\documentstyle[12pt,psfig]{article}
\begin{document}
\bibliographystyle{unsrt}

\begin{center}
{\Large{
\bf Order-Disorder Transitions and Melting in a Helical Polymer Crystal:
Molecular Dynamics Calculations of Poly(Ethylene Oxide) }}\\
\vspace{0.5cm}
M. Krishnan$^{+}$, S. Balasubramanian$^{+}$\footnote[1]{
Corresponding Author; email: bala@jncasr.ac.in} \\
$+$ Chemistry \& Physics of Materials Unit,\\
Jawaharlal Nehru Centre for Advanced Scientific Research,\\
Jakkur P.O., Bangalore 560 064, India\\
\vspace*{1.5cm}
{\bf Abstract}\\
\end{center}
Structural transitions and the melting behavior of crystalline poly(ethylene
oxide), (CH$_2$-CH$_2$-O)$_n$, (PEO) has been investigated using fully
atomistic, constant pressure-constant temperature (NPT) molecular dynamics (MD)
simulations. Melting of PEO proceeds in two stages; Order parameters reveal the
loss of interchain orientational correlations and of chain helicity in the
first and second stages of melting respectively. Sliding diffusion and
anisotropic reorientational dynamics of the polymer backbone are observed in
the solid state, in agreement with $^1$H NMR experiments.
\newpage

Relaxation processes in helical solid polymers such as poly(tetrafluoroethylene) (PTFE),
poly(ethylene oxide) (PEO), and polypropylene,
have been studied extensively over several decades~\cite{wunderlichreview,strobl_book}. 
Conformational transitions in
crystals of such polymers and of long chain molecules can lead to molecular
mobility in the solid state~\cite{first5}.  The coupling between intramolecular
order and intermolecular interactions in such systems can induce
structural transitions within the solid state, leading to a rich phase
diagram~\cite{frenkel2003,ishkiriyama}. 
 For instance, in PTFE, 
x-ray diffraction and calorimetry have shown the existence of 
 three solid phases at atmospheric pressure~\cite{muus}. 
Recent atomistic simulations of PTFE have
been able to reproduce the existence of these phases, and have provided
microscopic structural data that differentiates them~\cite{sprik}.

 The chains
in PEO, unlike those in PTFE, have a distorted helical conformation in the
crystalline phase.  The melting point of PEO is around 340K, with a dependence
on the molecular weight and on the morphology of the crystal~\cite{kovacs_fold}.
The calorimetric trace of low molecular weight PEO containing extended chains, exhibits 
a hump or a long tail at temperatures below the melting transition~\cite{buckley75}, 
which suggests that melting could involve stages of microscopic events.

Early studies on $^{1}$H NMR line widths of PEO had shown a sudden decrease
at 270K, much below its melting point~\cite{hikichi}.
Calorimetric~\cite{buckley75,lang1}, diffraction~\cite{saruyama_83}, and
spectroscopic~\cite{yoshihara} measurements as well have shown signatures of
premelting transitions, that lead to
molecular mobility in the solid state.
Our aim here is to provide complementary, microscopic data to identify such 
a transition in the 
crystalline phase of PEO.  In an effort to understand the structural
changes accompanying this transition, we have performed atomistic molecular
dynamics simulations of PEO as a function of temperature~\cite{md_details}.
Two transitions are observed -- one associated with interchain ordering, and
the other with intrachain ordering. The intermediate phase, although crystalline,
exhibits interesting translational and rotational dynamics.

Figure~1 presents the internal energy of the system, divided into
intermolecular and bonded intramolecular contributions as a function of
temperature.  The latter is a sum of contributions from stretch, bend, and
torsional interactions. Its dependence on temperature is monotonic,    
indicating the gradual population of defect states, 
while two clear discontinuities can be seen in the behavior of intermolecular energy.
This shows that the simulated PEO melts in two stages.
This is also evident from the evolution of the lattice parameters as a function
of temperature shown in Figure~1. The $c$ parameter, along which the
polymer chains are aligned,  shows a large decrease only at around 470K, i.e.,
during the second stage of melting. However, the $a$ and $b$ parameters exhibit
a behavior similar to the intermolecular energy, with two discontinuities, one
at 390K, and another at 480K, indicating again, the relationship between
intermolecular order and the transition at 390K.  These two stages of melting
are also seen in the variation of the volume of the system. Based on
the changes in both the internal energy and in the volume of the system, we
estimate the enthalpy change to be 3.0~kJ/mol of monomer for the
transition at 390K, and 6.2~kJ/mol of monomer for the second stage of
melting at 480K.  The sum of these values are in reasonable agreement with the
experimental value of 8.7~kJ/mol of monomer\cite{enthalpy_expt}.  We
have characterized the nature of these transitions using order 
parameters (Figure~1f) that
describe intermolecular orientations, as well as an intramolecular
characteristic, {\em viz.,} the helicity of the chains, both of which have been introduced
by Sprik {\em et al} earlier~\cite{sprik}.  The interchain
order parameter is obtained by studying the ordering in the orientation of the
bisector of methylene groups that belong to chains separated by distances of
the order of 10\AA.  The helical order parameter studies the angle between
vectors that bisect successive triplets of atoms in a polymer backbone.
Consistent with our observations earlier, we find that the interchain
orientational order collapses at the premelting transition, while the helical
or intramolecular ordering vanishes at the melting transition. Thus, it appears 
that although the interchain CH$_2$ orientational ordering is lost at around
390K, the chains retain their helicity and their orientation along the $c$-axis
through this premelting, solid-solid transition~\cite{lang1}.

The loss of interchain ordering, seemingly innocuous, is accompanied by subtle
conformational transitions in the polymer chains.  At room temperature, the
torsions around the C-C and the C-O bonds are in {\em gauche} ($g$) and {\em
trans} ($t$) states respectively, with the sequence, $ttg$ for successive bonds
in the chain producing the basic helical conformation. 
In the crystal, the C-O torsional angle is found to have a value of 
186.6$^o$$\pm$7.8$^o$~\cite{takahashi}. This width in the torsional distribution
manifests as a distortion in the helical conformation and is attributed to the
flexibility of the PEO chains and to strong intermolecular
interactions~\cite{takahashi}.  It is thus natural to ask if the observed loss
of ordering in interchain CH$_{2}$ orientations results in the reduction of this
distortion.  The distributions of C-O torsional angles obtained from our
simulations at different temperatures, are displayed in
Figure~2a~\cite{time_average}. We observe that torsional sub-states present near angles of 
150$^o$ and 210$^o$ that were presumably frozen by strong intermolecular
interactions, are populated at 310K and 380K, but become less likely at 390K and at
410K.  This implies that the polymer chains are less distorted
with increasing temperature within a narrow window of 20K around 400K.  Cross
sectional views of the crystal in the $ab$ plane shown in Figure~2b shows the 
distortion in the molecular conformation at 380K, while at 410K, a good fraction of the 
molecules exhibit a smoother, nearly circular pattern.  This
change can be attributed, among other
possibilities, to a reduction in the distortion in the helix, consistent with the 
results on C-O torsional distributions observed in Figure~2a. The simultaneous
emergence of this intramolecular, ``nearly ordering'' transition and the loss of
orientational order between chains, further validates the original observations
of Takahashi and Tadokoro~\cite{takahashi} that the conformational distortion
is brought about by interchain forces.  At temperatures higher than 410K, the
number of C-O torsions in the {\em gauche} state increases gradually, thus
decreasing the helicity of polymer chains. The chains approach a random coil
configuration (not shown), causing the $c$-parameter of the crystal unit cell to decrease.
The helical order parameter vanishes at around 480K, signifying the completion
of the melting process.

The structural changes through the two stages of melting bring forth
interesting behavior in the dynamics of the polymer chains. Specifically, the
loss of orientational ordering between chains at 390K and their adoption of 
smoother contours
allows them to slide past each other along
the direction of the $c$-axis in the crystal. In Figure~3, we show the
displacements of the centers of mass of the polymer chains along the $c$-axis,
and in the $ab$ plane at a few selected temperatures, across the two
transitions. We find that the displacement curve for motion along the $c$-axis
exhibits a positive slope at 390K, which is absent at 380K.  This shows
that the premelting transition is reasonably sharp.
As expected, the molecular displacements in the $ab$
plane is minimal up to even 410K. The lateral motion increases in magnitude
only much after the premelting transition. Our data on the sliding diffusion of 
chains is in good agreement with numerous experimental 
observations~\cite{olf_peterlin,hikosaka}. The anisotropy in molecular
displacements that sets in at 390K and its continuance up to the melting
transition at 480K is evident and explains the NMR and calorimetric 
observations~\cite{hikichi,lang1}.

We further probe the dynamics of the system through time correlation functions
(TCF) of the bisector ($\vec{R}$) of triplets of consecutive atoms along the
polymer backbone. In Figure~4, we display the normalized time correlation
functions of the projection of $\vec{R}$, on to the $c$-axis, and its residual
in the $ab$ plane.  We denote these as $C_c(t)=\langle R_c(0)R_c(t)\rangle$ and
$C_{ab}(t)= \langle \vec{R}_{ab}(0)\vec{R}_{ab}(t)\rangle$ respectively. At
310K and 380K, the function in the $ab$ plane does not relax for over hundreds
of picoseconds, as expected for a rigid solid.  The $C_{c}(t)$ function, on the 
other hand, relaxes
to a constant, non-zero value after about 1~ps at these temperatures. Its value
at long times is considerably smaller than that for the $C_{ab}(t)$ function,
indicating the greater spread of angles that the bisector can probe along the
$c$-axis. This enhanced angular freedom along the $c$-axis is presumably
due to the relative ease of the atoms of the chains to buckle back and forth,
rather than to make long amplitude vibrations along the $c$-axis.  This
observation is in good agreement with rms amplitudes of vibrations obtained
from x-ray diffraction studies in the isostructural polyoxymethylene
crystal~\cite{takahashi_tadokoro_1979}, and from the splittings of the
perpendicular bands observed in the infrared spectra of PEO at low
temperatures~\cite{yoshihara}.  At 390K, the $C_{ab}(t)$ function relaxes
slowly to a non-zero value, implying that the projection of the bisector vector
in the $ab$-plane is able to probe only a large portion of a circle in the
plane, and not the entire range of 360$^o$. At 410K and above, the function
decays to zero, a consequence of the rotation of the polymer chains about
their long axis. This rotation is facilitated by the formation of the
cylindrical contour of the chains, as discussed in Figure~2. A surprising
observation is that the  $C_{c}(t)$ function exhibits a slower decay with
increase in temperature between 410K and 480K. This apparently anomalous
observation can be rationalized if one considers that the formation of
{\em gauche} defects along the chain decreases the $c$ lattice parameter, and
increases the density of the system along this direction, thus slowing the
relaxation of the $C_{c}(t)$ function.  At 480K, the dynamics along the
$c$-axis and that in the $ab$ plane, are nearly identical, indicating the
attainment of isotropy in the system.

To summarize, large scale, atomistic MD simulations of PEO
crystal at more than 30 temperatures, have revealed interesting conformational, and intermolecular transitions
that profoundly influence the dynamics of the polymer chains. Melting of the
crystal takes place in two stages~\cite{superheat}. In the first stage, at
390K, we observe a discontinuous, solid-solid transition, much like first
order, with signatures in thermodynamical and structural quantities. Unlike
Sprik {\em et al}~\cite{sprik}, we observe a clear discontinuity in this first stage 
of the transition, probably due to the finer temperature resolution that
we were able to employ.  This premelting transition is associated with the loss
of interchain orientational ordering, and is accompanied by an interesting
change in the conformation of the chains -- over a narrow temperature range around 
400K, there is a small reduction in the helical distortion.
This transformation aids the
translational diffusion of chains along their long axes.  The second stage of
melting is completed at 480K, where the helicity of the chains
is entirely lost, resulting in randomly oriented chains. 
Consequently, the $c$ lattice parameter decreases
monotonically during this stage between 400K and 480K.  The ability of the
chains to rotate about and diffuse along their long axes is established at
390K. The TCF of the backbone vector along the long axis of the chains exhibits
a seemingly anomalous temperature dependence during the second stage of
melting, i.e.,  with increasing temperature, it relaxes slower. However, this
can be rationalized in view of the increase in density along the $c$-axis during this phase. 
The results reported here are comparable to experimental data on single crystals of
extended chain PEO. Calorimetric traces of such low molecular weight PEO have exhibited a hump
or a long tail at temperatures below the melting transition~\cite{buckley75}. 

Despite differences in the crystal structures and in the conformations of chains, 
the structural phase transitions observed for PEO here bears some similarities  
to those seen in PTFE by Sprik and coworkers~\cite{sprik}. 
It is to be noted however, that the high temperature phase of PTFE is columnar with
hexagonal symmetry~\cite{ungar_review}, while that in PEO continues to be a three
dimensional crystal.
The similarity of these phase
transformations suggest that the phenomena observed might be universal and could
be applicable to crystals of biologically relevant, small organic molecules
possessing helical conformations. 
The sliding diffusion~\cite{olf_peterlin,hikosaka} of polymer chains in
the orientationally disordered solid phase adds support to the mechanisms of
folding transitions observed in metastable forms of 
PEO~\cite{kovacs_fold,cheng_jpsb86} and of crystallization
kinetics in polymers~\cite{strobl_epje01}.

We thank Professor Biman Bagchi for helpful comments on an earlier version of this manuscript,
and the CSIR, India for partial financial support.

\newpage

\clearpage
\newpage
\begin{figure} 
\centerline{\psfig{figure=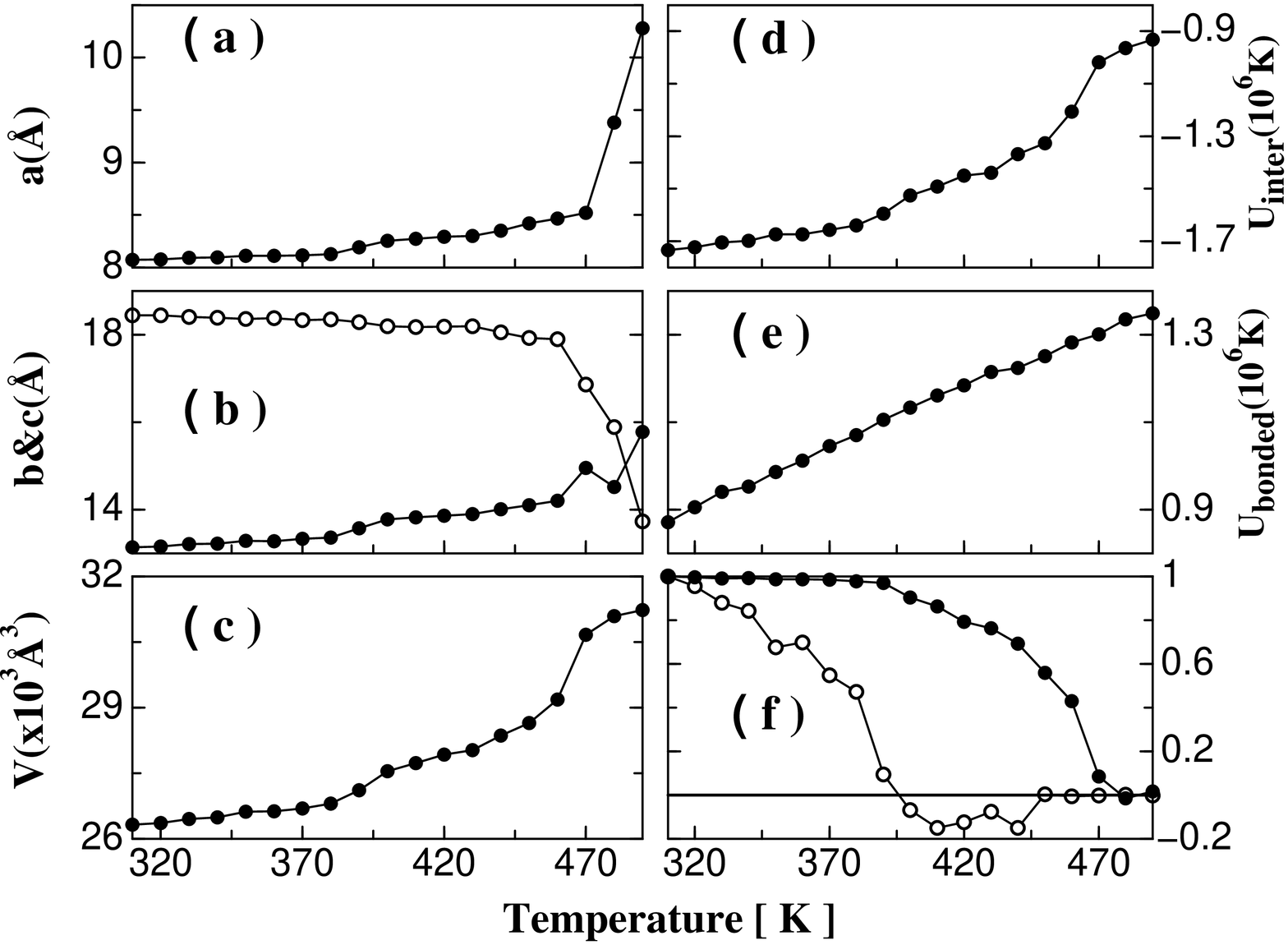,height=4.8in,angle=270}} 
\vspace*{2.0cm}
\caption{
Lattice parameters of unit cell (a) $a$ (b) $b$ (filled
circles) and $c$ (open circles), and (c) Volume of the simulation cell (d)
Intermolecular energy (e) Bonded intramolecular energy, (f) Interchain
orientational order parameter(open circles) and helical order parameter
(filled circles) against temperature.  The order parameters
are normalized with respect to their values at 310K.}
\label{F_fig1} 
\end{figure}

\newpage
\begin{figure} 
\centerline{\psfig{figure=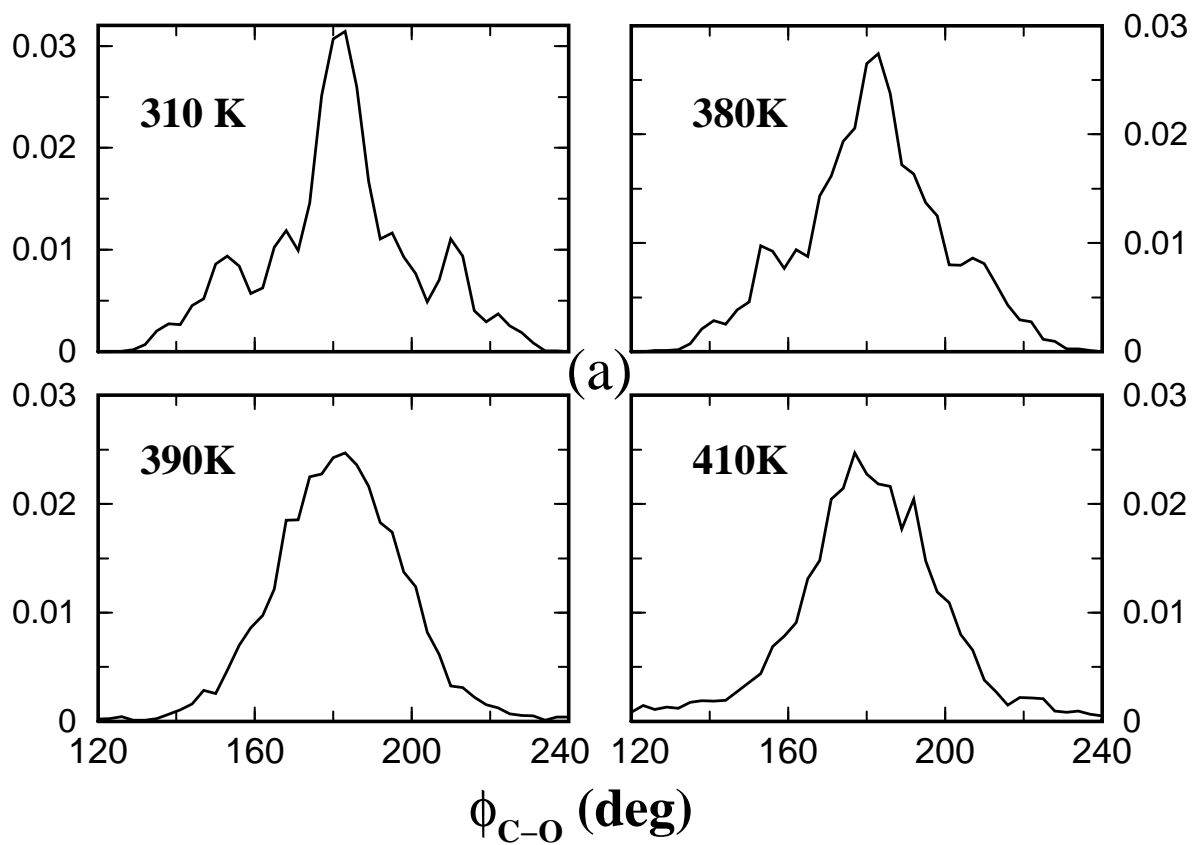,height=4.4in,angle=270}} 
\label{F_fig2a}
\centerline{\psfig{figure=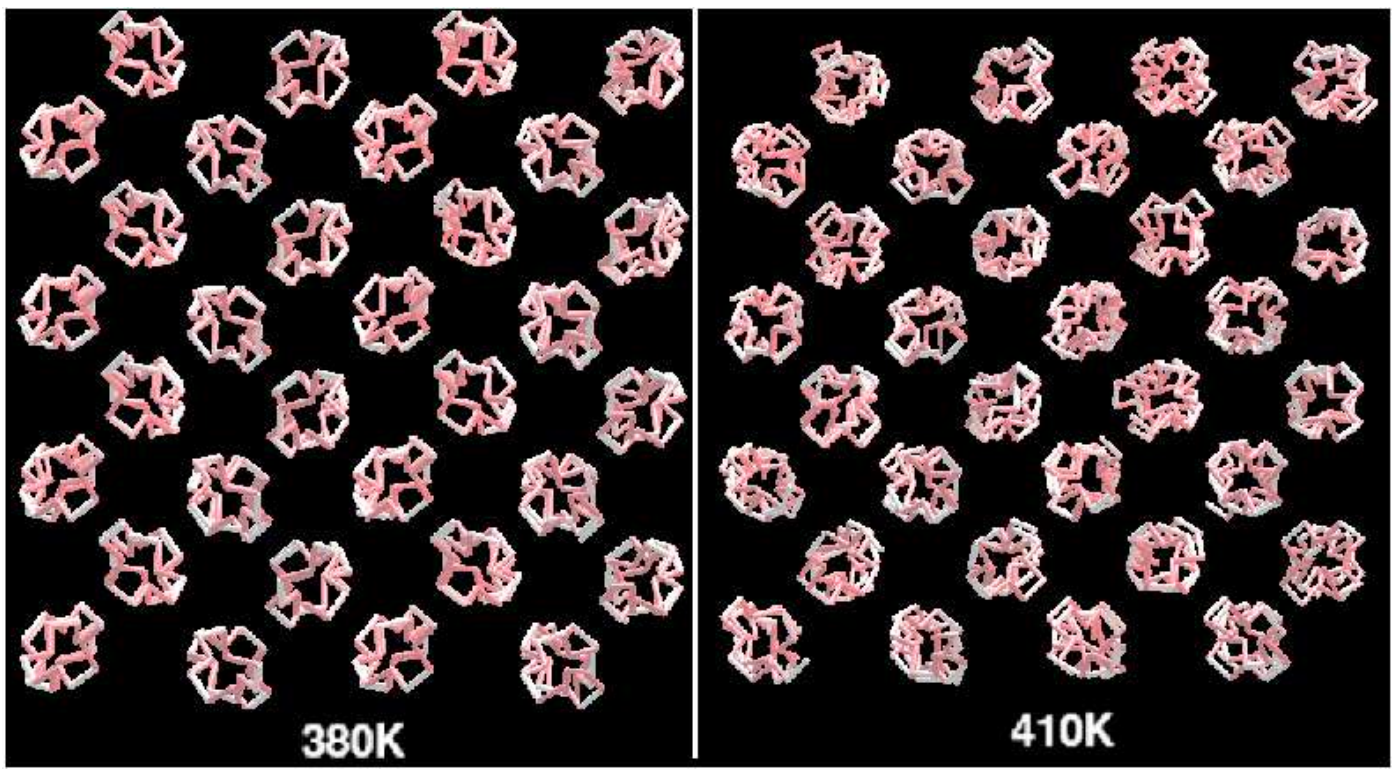,height=4.0in,angle=270}} 
\vspace*{0.3cm}
\caption{
(a)Distribution of C-O torsional angles, obtained from time averaged 
configurations~\cite{time_average} for temperatures indicated.
(b)View through $c$-axis of time averaged configurations at two temperatures.}
\end{figure}

\newpage
\begin{figure} 
\centerline{\psfig{figure=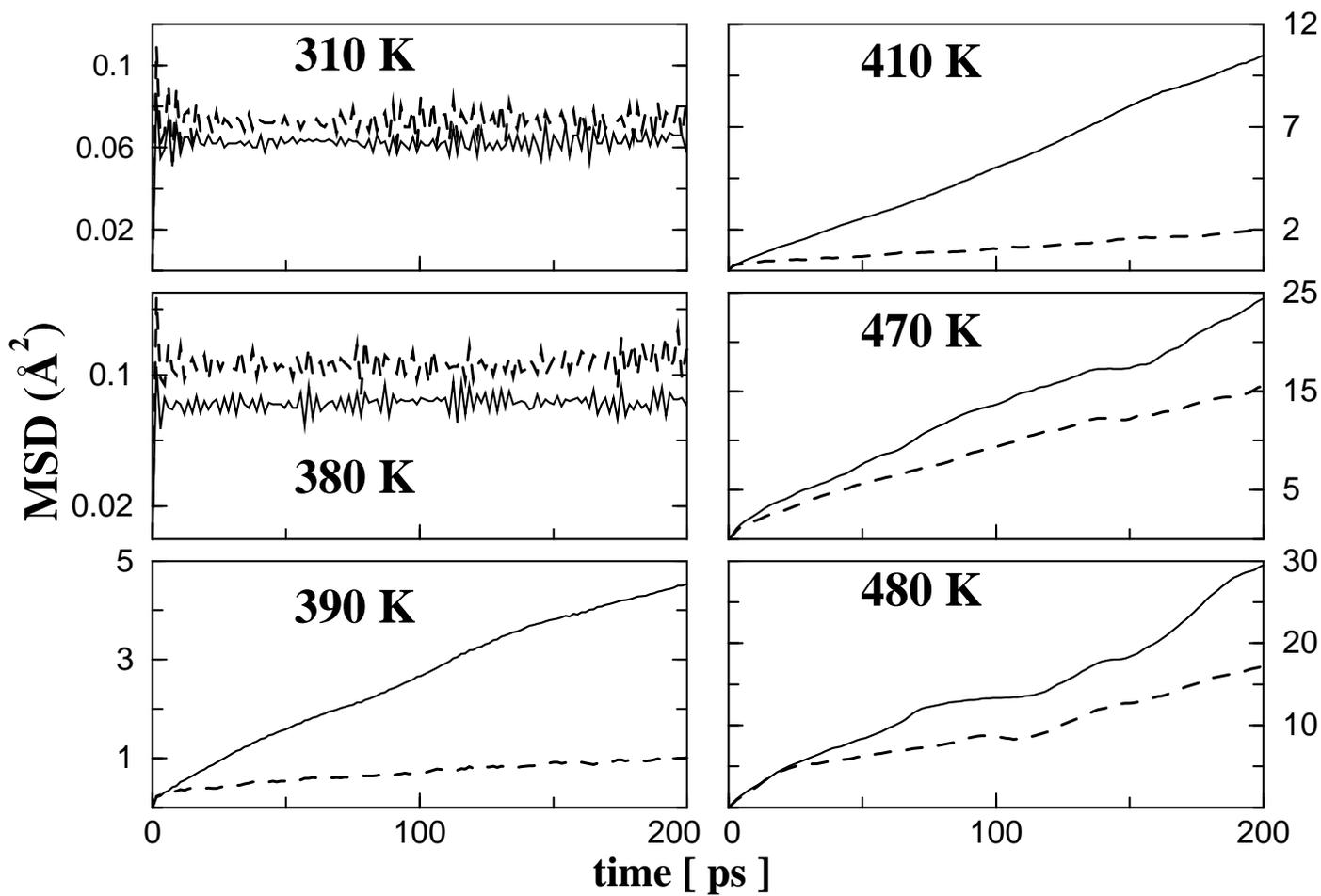,height=4.7in,angle=270}} 
\vspace*{2.0cm}
\caption{
Mean square displacement of centers of mass of PEO chains
along the $c$ axis (solid lines) and in the $ab$-plane (dashed lines) for
temperatures indicated.}
\label{F_fig3} 
\end{figure}

\newpage
\begin{figure} 
\centerline{\psfig{figure=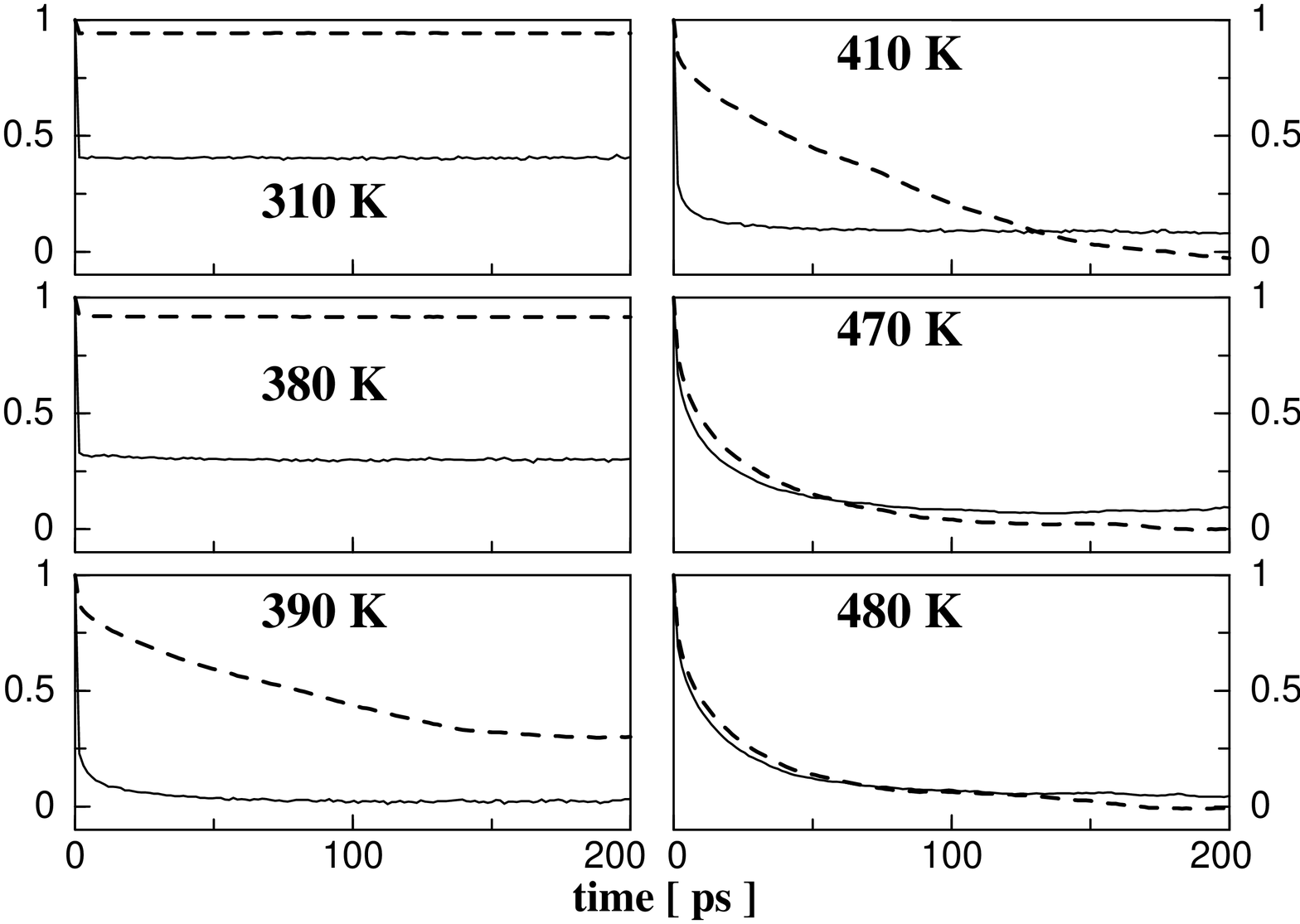,height=4.7in,angle=270}} 
\vspace*{2.0cm}
\caption{
Normalized rotational time correlation functions,
C$_{c}(t)$ (solid lines) and C$_{ab}(t)$ (dashed lines) at different
temperatures. }
\label{F_fig4} 
\end{figure}

\begin{thebibliography}{99}
\bibitem{time_average} The conformational transition at 390K can be observed by
using time averaged configurations that filters thermal vibrations. The
analyses reported in Figure~2 are obtained over a short run of length 5.4~ps
with a time resolution of 27~fs, so as to minimize the effects of chain
diffusion.
\bibitem{pitch}As the polymer chains in our simulation were connected
head-to-tail to remove end effects, conformational excitations which produce a
helical pitch that is a non-integral divisor of the simulation cell along $c$
axis are forbidden.
\end{thebibliography}
\end{document}